\documentclass[aps,amsmath,twocolumn,amssymb,floatfixng,showpacs,superscriptaddress,footinbib]{revtex4-1}
\usepackage[T1]{fontenc}
\usepackage[utf8]{inputenc}

\pdfoutput=1
\usepackage[dvips]{graphics}
\usepackage{bm}
\usepackage{float}
\usepackage{epsfig}
\usepackage{enumerate}
\usepackage{subfigure}
\usepackage{color}
\usepackage{braket}
\usepackage{graphicx}
\usepackage{float}
\usepackage[colorlinks=true,linktoc=page,linkcolor=red,citecolor=blue,urlcolor=magenta]{hyperref}
\usepackage{amstext}
\usepackage{lipsum}
\usepackage{mathtools}
\usepackage{afterpage}
\usepackage{epstopdf} 
\usepackage{filecontents}
\usepackage{soul}

\newcommand{\myvec}[1]{\textbf{#1}}

\newcommand\bea{\begin{eqnarray}}
\newcommand\eea{\end{eqnarray}}
\newcommand\beq{\begin{equation}}  
\newcommand\eeq{\end{equation}}

\usepackage[normalem]{ulem}
\usepackage{sidecap,tikz}
\DeclareRobustCommand{\orcidicon}{\hspace{-1.0mm}
	\begin{tikzpicture}
	\draw[lime, fill=lime] (0.0,0.0) 
	circle [radius=0.15] 
	node[white] {{\fontfamily{qag}\selectfont \tiny \,ID}};
	\draw[white, fill=white] (-0.0525,0.095) 
	circle [radius=0.007];
	\end{tikzpicture}
	\hspace{-3.0mm}
}
\foreach \x in {A, ..., Z}{\expandafter\xdef\csname orcid\x\endcsname{\noexpand\href{https://orcid.org/\csname orcidauthor\x\endcsname}
		{\noexpand\orcidicon}}
}

\begin{document}
\title{Multiple higher-order topological
phases with even and odd pairs of zero-energy corner modes in a $C_3$ symmetry broken model}
\author{Sudarshan Saha}\email{sudarshan@iopb.res.in}\affiliation{Institute of Physics, Bhubaneswar- 751005, Odhisa, India}
\affiliation{Homi Bhabha National Institute, Mumbai - 400 094, Maharashtra, India}
\author{Tanay Nag\orcidB{}}
\email{tanay.nag@physics.uu.se}
\affiliation{Department of Physics and Astronomy, Uppsala University, Box 516, 75120 Uppsala, Sweden}
\author{Saptarshi Mandal\orcidC{}}\email{saptarshi@iopb.res.in}\affiliation{Institute of Physics, Bhubaneswar- 751005, Odhisa, India}
\affiliation{Homi Bhabha National Institute, Mumbai - 400 094, Maharashtra, India}

\begin{abstract}
Here we study emergent higher-order topological (HOTI) phases in the extended  Haldane model  without $C_3$ symmetry. For inversion symmetric case, the QSHI and QAHI phases can embed the HOTI phases while remaining QASHI phase does not yields any HOTI phases. Remarkably, four-fold degeneracy of zero-energy corner states can be reduced to two-fold under the application (withdrawn) of sub-lattice mass (Zeeman field) term. The sub-lattice mass and Zeeman field terms compete with each other to pin down the two mid-gap states at zero-energy. Interestingly, the bulk-polarization can  topologically characterize the second-order topological insulator phase with the mid-gap corner modes irrespective of their energies  as long as inversion symmetry is preserved.  Our study indicates that a hybrid symmetry can in principle protect the second-order topological insulator phases, however, spin-spectrum gap has to be essentially finite there.
\end{abstract}

\date{\today}

\maketitle


\underline{\textit{Introduction:}}

The higher-order topological insulator (HOTI) phases \cite{Benalcazar17,schindler2018higher}  have drawn a considerable research endeavour in the field of topological systems \cite{Haldane88,Onoda03,Kane05a,Kane05b,Bernevig06,Klaus86,Maciejko11,Hasan10,Liu16a} due to their intriguing bulk-boundary correspondence. There exist various symmetries such as, reflection,  inversion, rotational and time-reversal symmetries which are becoming instrumental in engineering the HOTI phases \cite{Zhida17,Langbehn17,Khalaf18,Miert18,Dumitru19}. For example, the second-order topological insulator (SOTI) phase in two-dimensional (2D) quantum spin-Hall insulator and topological superconductor host $0$-dimensional (0D) electronic and Majorana corner modes, respectively, which are characterized by various topological invariant such as, quantized quadrupole moment, edge and bulk polarizations, Wannier centers(WCs) \cite{Liu18,Wang18,Volpez19,Roy20,Ghorashi20,Wu20,Ghosh21c,Yang21,QWang18}. Importantly, HOTI phases have been theoretically predicted and experimentally realized in honeycomb  \cite{Ezawa18,Liu19,Mizoguchi19,Lee20,Xue21,Zangeneh19}, Kagome \cite{Ezawa18b,van2020topological,el2019corner,li2020higher,xue2019acoustic,ni2019observation} lattices, and Kekulé-distorted Haldane model \cite{Bunney22} extending their  presence beyond the realm of  square or cubic lattices \cite{Franca18,Nag19,Seshadri19,Bomantara19,Huang20,Haiping20,Nag21,Ghosh21,Ghosh21b,Zhu21,RXZhang21,Ghosh22a,Ghorashi20,Wang20,Chen19,Xie19,imhof2018topolectrical}.
The higher-order corner modes are predicted under $C_3$ symmetry breaking in the modified Haldane model that hosts first-order quantum anomalous Hall insulator (QAHI) phases otherwise  \cite{Wang21}.


The time-reversal symmetry (TRS) broken quantum spin Hall insulator (QSHI) \cite{Qiao10,Yang11,Liu08,Li13}, based on graphene, is 
not extensively studied so far in the context of SOTI phases. The Kane-Mele model, preserving inversion symmetry (IS), with a magnetic field can host such HOTI phases in honeycomb lattice \cite{Yafei20}. On the other hand, the IS-protected HO analogue of QAHI can be engineered only in absence of sub-lattice mass \cite{Wang21}.  Motivated by the above studies, we here seek the answers for the following question: Can we engineer mid-gap zero-energy HOTI states using the interplay of sub-lattice mass, Zeeman exchange field and SOC interaction?
To be precise, the HO analogue of QAHI, QSHI and quantum anomalous spin  Hall insulator (QASHI) phases, as already observed in the extended  Haldane model \cite{Saha21}, is the main focus of the present work. Our aim is to analyze the importance of spin-spectrum gap as well.

\begin{figure}[!htb]
    \centering
	\includegraphics[width=0.48\textwidth]{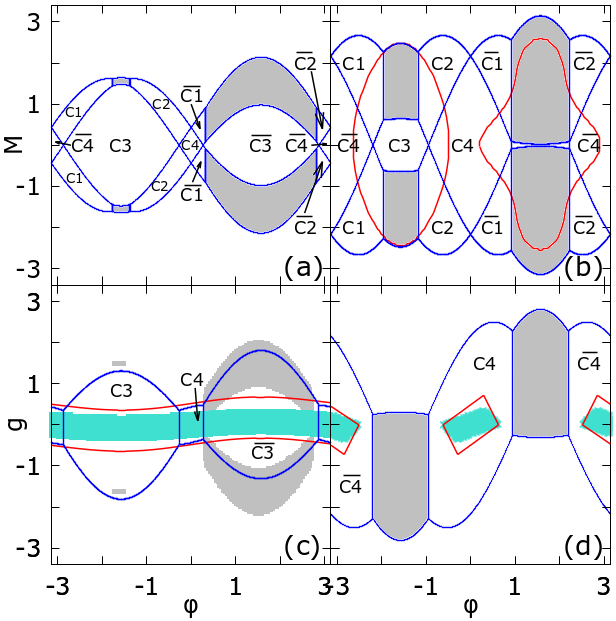}
\caption{ (a) and (b) show phase diagram in $M-\phi$ plane for $V_{\rm SO} = 0.5$ and $2.5$, respectively with  $g=-0.33$. Similarly (c) and (d) show phase diagram in $g-\phi$ plane for $V_{\rm SO} = 0.5$ and $2.5$, respectively with  $g=-0.33, M=0$. The detailed discussion is provided in the main text (second page and left column).}	
	
	\label{fig:1}
\end{figure}

\begin{figure}[!htb]
	\includegraphics[width=0.48\textwidth]{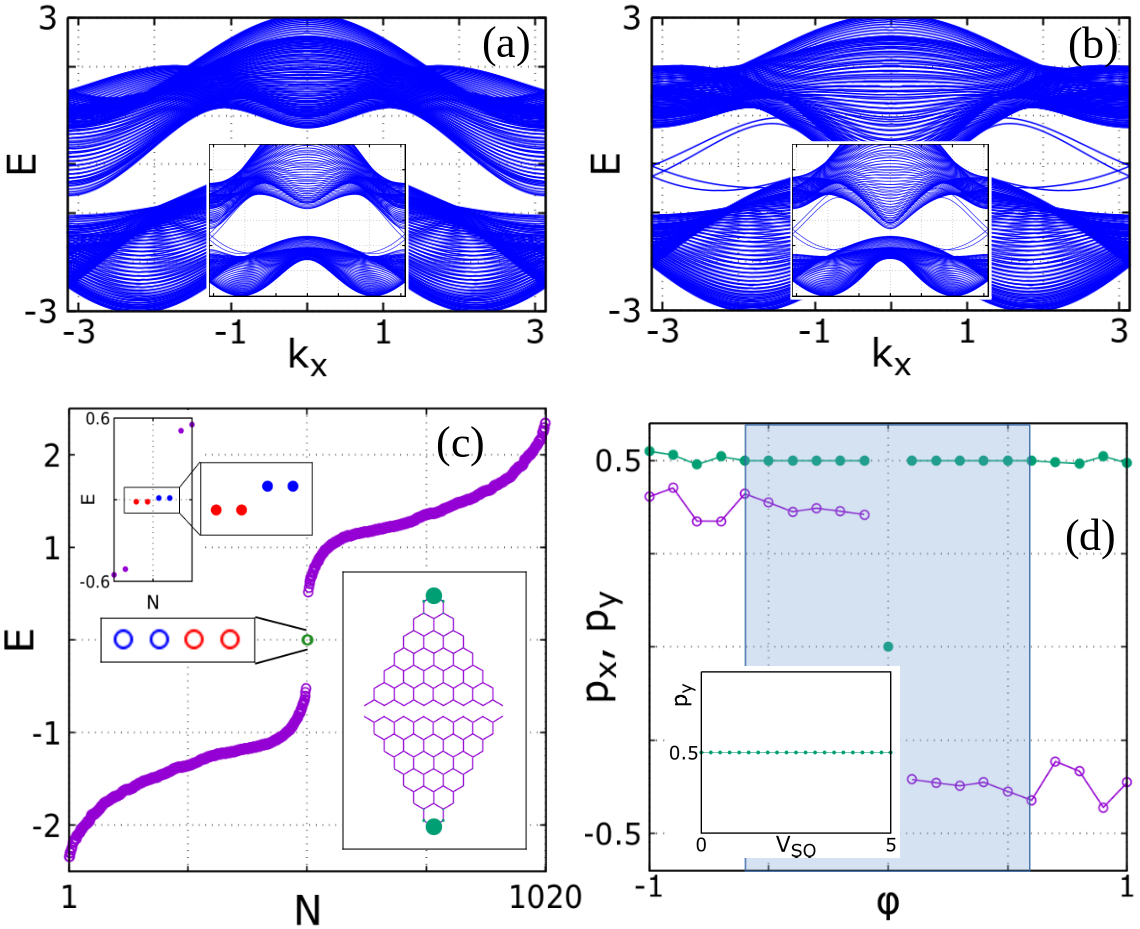}
\caption{The ribbon geometry dispersion is shown for $\eta=0.25$ along cut 2 and cut 1 in (a) and (b) respectively. In the inset of (a) and (b) dispersion for $\eta=0.75$ is shown. $E$(energy) vs $N$ for cut 2 nano-disc geometry is shown in (c) and  the inset  shows localization of mid gap states at the corner of nano-disc for $g=-0.023$ ($g=-0.01$, upper inset). In (d) the green solid-circle (pink empty-circle) shows variation of $p_y$($p_x$) for $g=-0.023$, inset shows variation with $V_{SO}$. Here we have considered $M=0$, $\phi = 0.1$, and $V_{\rm SO} = 2.5$ throughout so that we are inside SOTI phase, embedded in the underlying QSHI phase, as shown in Fig.~\ref{fig:1} (d).}	
	
\label{fig:2}
\end{figure}

\begin{figure}[!htb]
	\includegraphics[width=0.48\textwidth]{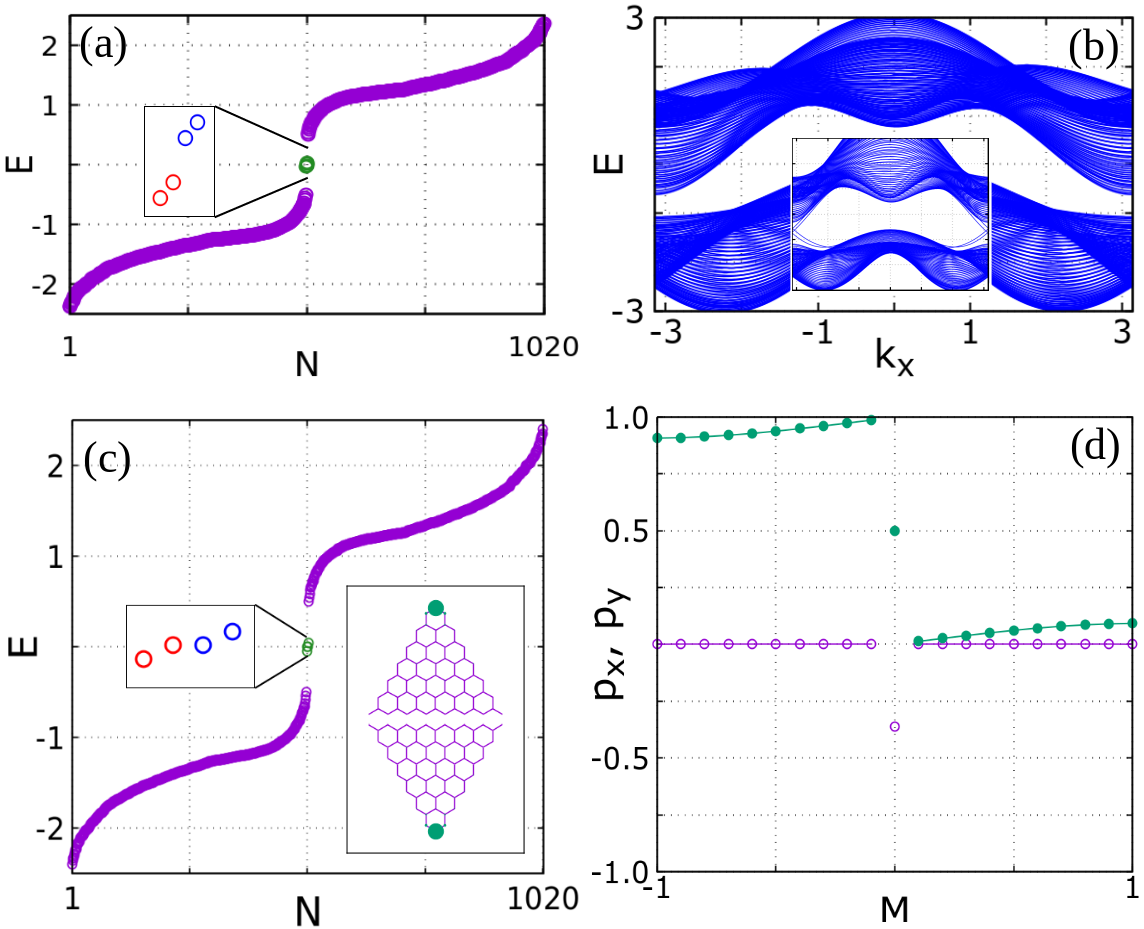}
	\caption{(a) The four non-degenerate mid-gap states at finite energies are observed in $E$ vs $N$ for nano-disc with $M= -0.01$.	The gapped band structure [energy dispersion with two-fold degenerate zero-energy and two non-degenerate finite energy mid-gap corner states], considering ribbon [nano-disc] geometry for cut 2, is shown in (b) [(c)] where $M=-0.037$. The gapless  edge states exist for cut 1 as shown in the inset of (b). (d) The bulk-polarization $p_y$ deviates from $0.5$ for $|M|\ne 0$ even though zero-energy corner states are present.  We considered $\phi = 0.1$, $V_{\rm SO} = 2.5$, and $g = 0.0$ such that we are inside the SOTI phase as shown in Fig.~\ref{fig:1} (d).}
	\label{fig:3}
\end{figure}

\begin{figure}[!htb]
    \includegraphics[width=0.48\textwidth]{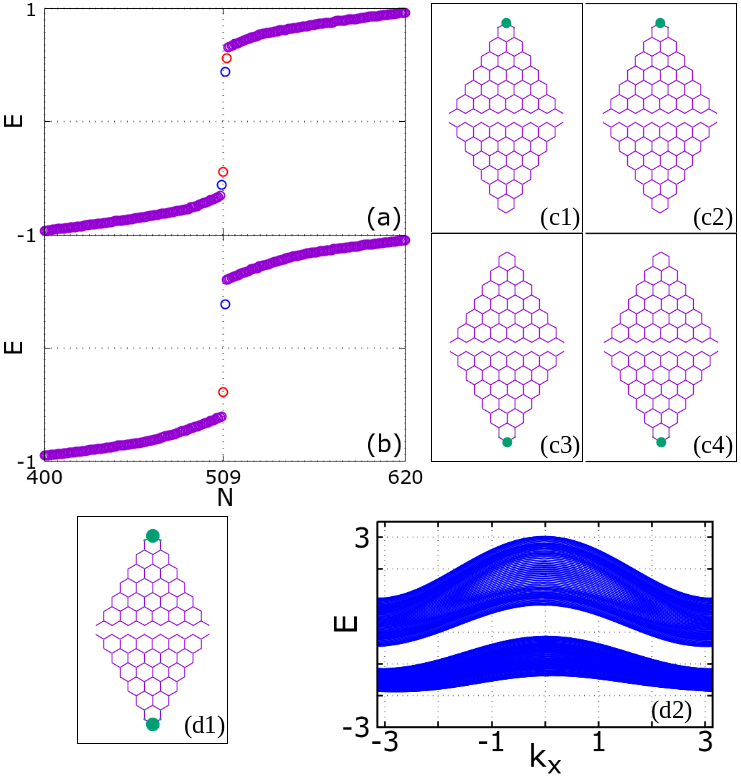}
	\caption{We depict the energy dispersion under nano-disc geometry with cut 2 for $g=-0.06$, and $-0.11$ in (a) and (b) respectively, considering $M = 0.5$, $\phi = 0.1$, and $ V_{\rm SO} = 0.0$.  The two mid-gap states move towards the zero-energy and the other two states are pushed towards the bulk bands when $|g|$ increases. We demonstrate the spatial localization of the four mid-gap
    modes, denoted by top most red, blue, red, and bottom most blue circles in (a), in (c1), (c2), (c3) and (c4), respectively. 
    Two mid-gap states with positive energy are localized at $(0,+L_y)$ while the remaining two with negative energy are at  $(0,-L_y)$.  In (d1), we show the combined spatial localization for the two mid-gap states at finite-energy in (b).  The momentum space band structure for (b) is displayed in  (d2).}
	\label{fig:4}
\end{figure}

\begin{figure}[!htb]
    \includegraphics[width=0.48\textwidth]{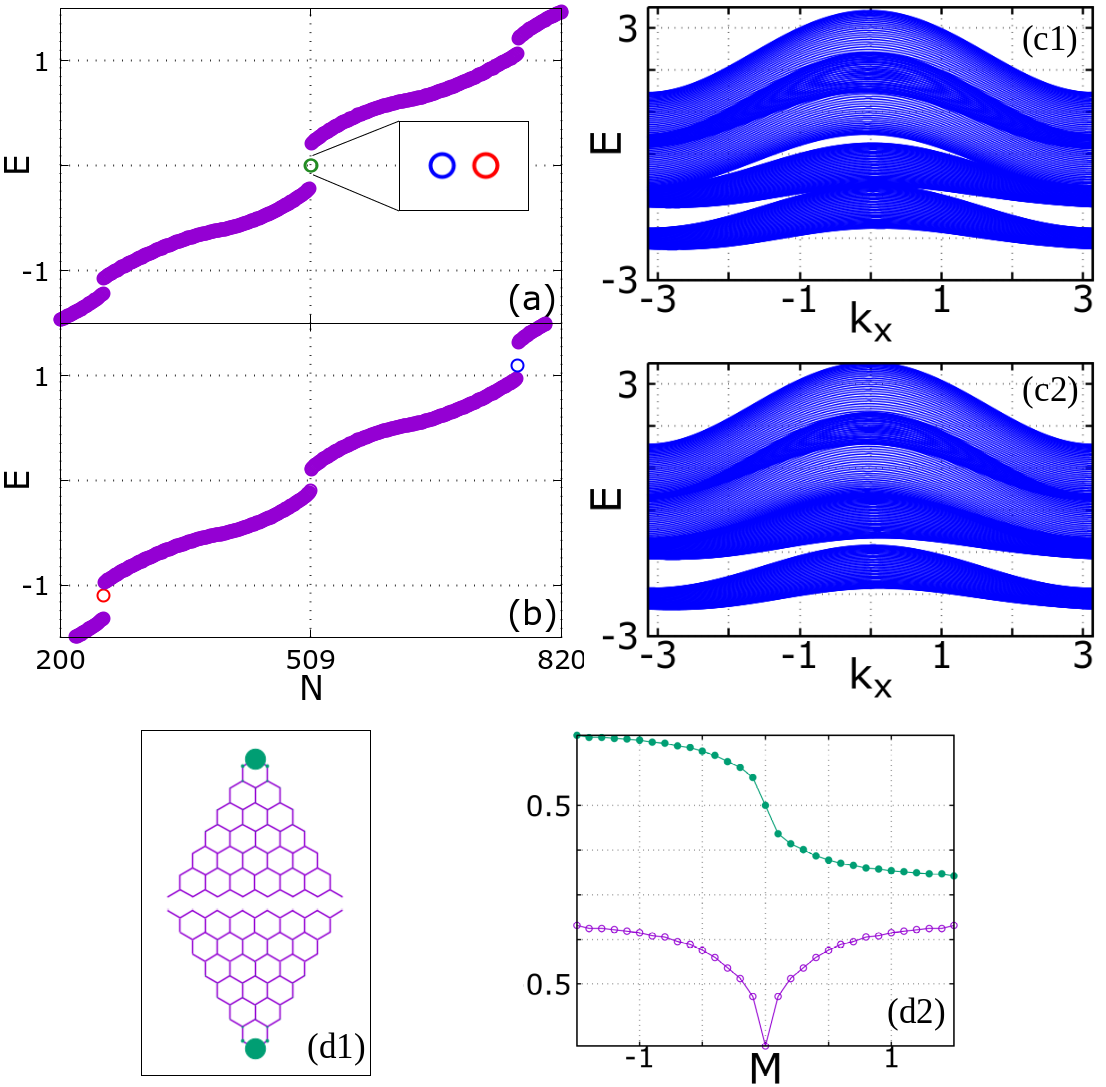}
	\caption{In continuation to Fig.~\ref{fig:4}, we show the energy dispersion under nano-disc geometry with cut 2 for $g=-0.5$ and  $-0.6$ in (a) and (b) respectively, considering $M = 0.5$, $\phi = 0.1$, and $ V_{\rm SO} = 0.0$. The two mid-gap states in Fig.~\ref{fig:4} (b) can be pinned down at zero (finite but substantially high) energy when $|M|=|g|$ ($|M|<|g|$). The ribbon geometry band structure for (a) and (b) are  shown in (c1) and (c2), respectively. In (d1),  we show the combined spatial localization for the two mid-gap states at zero-energy in (a).   In (d2), we illustrate the  variation of $p_x$ and $p_y$ as a function of $M$ with $|M|=|g|$.}
	\label{fig:5}
\end{figure}

\begin{figure}[!htb]
	\includegraphics[width=0.48\textwidth]{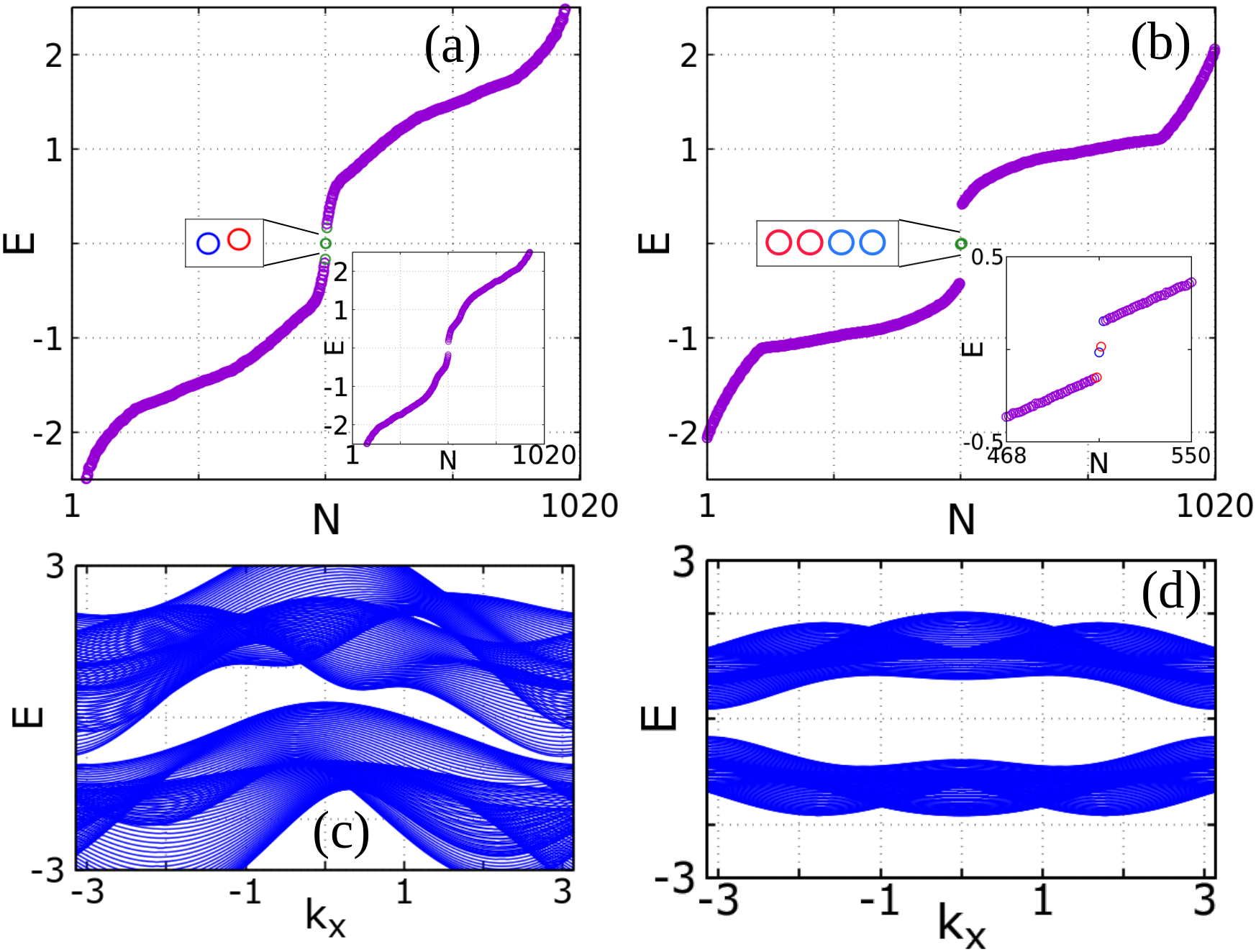}
	\caption{
   We show the mid-gap corner modes with  $(M,V_{\rm SO},g,\phi,\eta)=(-0.6,2.5,-0.33,-0.1,0.25)$ in (a), and $(0.0, 0.5, 0.074,1.5, 0.25)$ in (b) such that SOTI phases are chosen from the underlying QSHI, and QAHI phases in Fig.~\ref{fig:1} (b) and (c), respectively. 
   We can  pin down two [four] mid-gap states at finite- [zero-] energy in (a) [(b)] for $M\ne 0$ [$M=0$].  The energy dispersion without the mid-gap states is depicted as the inset in (a) for  $(1.2,2.5,-0.33,-0.1,0.25)$. 
    The energy dispersion with two mid-gap states is shown as the inset  in (b) for $(-0.4,0.5, -0.33, -1.57,0.25)$ chosen from the underlying QAHI phase in Fig.~\ref{fig:1} (a). 
   We illustrate the gapped band structure under ribbon geometry in (c) and (d) corresponding to (a) and (b), respectively. 
       }
	\label{fig:6}
\end{figure}

Considering $C_3$ symmetry and TRS broken extended Haldane model, we find that an interior part of QSHI and QAHI phases can only host mid-gap corner states while the  QASHI phase does not support any SOTI  phase in inversion IS case (see Fig.~\ref{fig:1}).The SOTI phase, hosting four zero-energy corner modes, is characterized by the bulk-dipole moment only when IS is preserved (see Fig.~\ref{fig:2}). In the presence (absence) of SOC interaction (Zeeman field), the earlier four-fold degeneracy of zero-energy corner states can be reduced to two-fold degeneracy when the IS is broken by a finite sub-lattice mass even though the bulk-polarization fails to detect the later phase (see Fig.~\ref{fig:3}). In the absence of SOC interaction, the sub-lattice mass and Zeeman field terms compete with each other to pin down two of the four mid-gap states at zero-energy (see Figs.~\ref{fig:4} and \ref{fig:5}). We also find an instance when four mid-gap corner modes exist at zero-energy for the SOTI phase that is embedded in the QAHI phase (see Fig.~\ref{fig:6}). The interplay between finite Zeeman exchange field, sub-lattice mass, and SOC interaction to tune the mid-gap corner modes can be qualitatively understood from the effective energy gap criterion. We find a hybrid symmetry that can protect the SOTI phases, however, the finite spin-spectrum gap is an essential requirement to observe such SOTI phases. 

\underline{\textit{Model Hamiltonian:}}
We start with the $C_3$ symmetry broken extended Haldane model as follows \cite{Wang21,Saha21,Haldane88,Kane05a}
\begin{eqnarray}
\label{main:ham}
	H = && -\sum_{\langle ij \rangle} t_1^{ij} c^{\dagger}_i c_j + \sum_{\langle\langle ij \rangle\rangle} t_2^{ij} e^{i \phi_{ij}} c^{\dagger}_i c_{j} + M \sum_i c^{\dagger}_i \sigma_z c_i  \nonumber \\
	&& + ~ \frac{i V_{\rm so}}{\sqrt{3}} \sum_{\langle\langle ij\rangle\rangle}e^{i \phi_{ij}} \nu_{ij} c^{\dagger}_i \sigma_z c_j + g \sum_i c^{\dagger}_i \tau_z c_i
\end{eqnarray}
where $c_i(c^{\dagger}_i)$ indicates the  creation (annihilation) operator
with ${\bm \sigma} \in (A,B)$ and ${\bm \tau} \in (\uparrow,\downarrow)$ representing the orbital and spin degrees of freedoms. The spin-independent 
nearest neighbour (NN) [next nearest neighbour (NNN)] anisotropic hoppings are represented by $t_1=(\eta \: \! t_{1},\eta \: \! t_{1}, t_{1})$  [$t_2=(t_{2}, t_{2}, \eta \: \! \! t_{2})$] along $\delta_1=(-1/2,1/2\sqrt{3})$, $\delta_2=(1/2,1/2\sqrt{3})$, and $\delta_3=(0,-1/\sqrt{3})$ [$a_1=(-1/2,\sqrt{3}/2)$, $a_2=(1/2,\sqrt{3}/2)$, and $a_3=(1,0)$]. The factor $|\eta| \ne 1$ (with $|\eta| < 1$) is  responsible for the breaking of $C_3$ symmetry allowing strong  and weak bonds of strengths $t_{1,2}$ and  $\eta \: \! t_{1,2}$, respectively. The SOC interaction of strength $V_{\rm SO}$ corresponds to the spin-dependent NNN hopping. The phase factor $e^{i\phi_{ij}}$ designates the staggered magnetic flux.
Notice that we consider uniform strength of SOC interaction for simplicity. $M$ ($g$) represents the IS breaking sub-lattice mass term  (TRS breaking magnetic field acting on spin-degrees of freedom). The factor $\nu_{ij}=(\myvec{d}_{ij}^1\times\myvec{d}_{ij}^2)_z $ contains  unit vectors $\myvec{d}_{ij}^{1,2}$ along the two bonds the electron traverses going from site $j$ to $i$ \cite{Saha21,Kane05a}. The momentum space Hamiltonian,  obtained from Eq.~(\ref{main:ham}), in the basis $(c_{A\uparrow}, c_{A\downarrow}, c_{B\uparrow}, c_{B\downarrow})$ is given by $H({\bm k},\eta)= \sum_{i=0}^5 n_i  ~ \Gamma_i $ with $\Gamma_i=\sigma_{i} \otimes \tau_0$ for $i=1,2,3$, $\Gamma_{4}=\sigma_{3} \otimes \tau_3$, $\Gamma_{5}=\sigma_{0} \otimes \tau_3$ and $\Gamma_{0}=\sigma_{0} \otimes \tau_0$. The components $n_i$ are given by:  
$n_0= 2 t_2 f({\bm k},\eta) \cos \phi $, $n_1=-t_1 (1 + 2 h({\bm k},\eta)   )$ ,
$n_2= -2 \eta t_1  \sin \frac{\sqrt{3}k_y}{2} \cos \frac{k_x}{2}$,
$n_3= M - 2 t_2 ~g({\bm k},\eta) \sin \phi $,
$n_4= \frac{V_{\rm so}}{3} g({\bm k},1) \cos \phi $, 
$n_5=g + \frac{V_{\rm so}}{3} f({\bm k},1) \sin \phi$, with $f({\bm k},\eta)= 2 \cos \frac{\sqrt{3}k_y}{2} \cos \frac{k_x}{2} + \eta\cos k_x$,
$g({\bm k},\eta)= 2 \cos \frac{\sqrt{3}k_y}{2} \sin \frac{k_x}{2} + \eta \sin k_x $,
$h({\bm k},\eta)= \eta \cos \frac{\sqrt{3}k_y}{2} \cos \frac{k_x}{2}$.


To this end, we consider $\eta=1$ to investigate  various FOTI phases before exploring their second-order counterparts. The FOTI phases for a TR symmetry broken system is characterized by spin-Chern number \cite{spin-chern2,Prodan_2010} which is numerically computed with Fukui method \cite{Saha21}. The Zeeman field $g$ destroys the $\phi \rightarrow -\phi$ symmetry by shrinking or expanding different phases as compared to $g=0$ case where the  eight-fold FOTI phases can symmetrically appear under $\phi \rightarrow -\phi$ \cite{Saha21}.  The size of QSHI (QAHI) phases increases (reduces) referring to the fact that magnetic field stabilizes QSHI phases (see Figs.~\ref{fig:1} (a), (b)). 
The SOC is also found to exhibit a similar tendency to  suppress the QAHI phase when its strength increases. The spin-spectrum gap depicted by gray-shaded region, computed for $\eta=1$,  follows the FOTI phase boundaries marked by blue lines. Interestingly, the spin-spectrum gap for $\eta=0.25$ demarcated by red lines is significantly different from its profile at $\eta=1$. The SOC strength has a severe impact on the spin-spectrum gap as it is always finite in Fig.~\ref{fig:1} (a)  while it can be vanishingly small within a certain region in Fig.~\ref{fig:1} (b). The turquoise-colored region in Figs. ~\ref{fig:1} (c) and (d) shows when the mid-gap corner states appear.
 The spin-spectrum gap plays very crucial role as far as the emergence of SOTI phases is concerned.
The SOTI phases can in principle be embedded in QAHI and QSHI (QSHI) for $V_{\rm SO}=0.5$ ($V_{\rm SO}=2.5$) as mentioned below while describing Figs.~\ref{fig:1} (c) and (d).


The gap vanishes at the Dirac points over the boundaries between  two FOTI phases or non-topological and FOTI phases. For $0.5<|\eta|<1$, two Dirac points appear at
$\myvec{K}^{\eta}_{\pm} = ( \pm 2\theta,  -2\sqrt{3}\theta )$ with $\theta={\rm arctan}(\sqrt{4 \eta^2 -1})$.
The phase boundaries  can be obtained when the gap vanishes for $H(\myvec{K}^{\eta}_{\pm},\eta)$ whose 
eigenvalues are given by $\lambda^{\pm}_1=n^{\pm}_3-n^{\pm}_4-n^{\pm}_5$, 
$\lambda^{\pm}_2=-n^{\pm}_3+n^{\pm}_4-n^{\pm}_5$, $\lambda^{\pm}_3=-n^{\pm}_3-n^{\pm}_4+n^{\pm}_5$, and 
$\lambda^{\pm}_4=n^{\pm}_3+n^{\pm}_4+n^{\pm}_5$ with 
$(n^{+}_3, n^{+}_4, n^{+}_5)= (M + 6 t_2 \sin \phi \sin u, -(2V_{\rm so}/3) \cos \phi \sin u (1-\cos u), g+ (2V_{\rm so}/3) \sin \phi [(\cos u + 1/2)^2 -3/4] )$, and $(n^{-}_3, n^{-}_4, n^{-}_5)= (M + 4 t_2 \sin \phi \sin u, -(2V_{\rm so}/3) \cos \phi \sin u, g- (V_{\rm so}/3) \sin \phi$),
$\cos u= -1/2\eta$. Note that when $\eta=0.5$ ($-0.5$), two Dirac points reside at the ${\bf M}$ (${\bf \Gamma}$)-points \cite{Wang21}.   We are interested in the SOTI  for $0<\eta<0.5$, for which the Dirac points are gapped out leading to an intriguing phase diagram. Considering $\cos u =0,\sin u =1$ for the above case, the effective bulk band-gap  $\Delta^{\pm}_{ij}=\lambda^{\pm}_i-\lambda^{\pm}_j$ can acquire various forms such as $(\Delta^{-}_{12},\Delta^{+}_{14})=(2M+8 t_2 \sin \phi -(4 V_{\rm so}/3) \cos \phi,-2 g+(2 V_{\rm so}/3) \sin \phi-(4 V_{\rm so}/3) \cos \phi)$. Therefore, the boundaries demarcating the FOTI phases become non-trivially modified for the SOTI phases that might be bounded by the zeros  of the above effective band-gaps. This can qualitatively explain the observation of SOTI phases in the \emph{interior} part of the underlying QSHI phase as shown in Figs.~\ref{fig:1} (c) and (d). Notice that corner states in modified Haldane model can be observed even for $M\ne 0$ inside the FOT Chern insulator phase \cite{Wang21}.

The SOTI phase is characterized by bulk-dipole moment $p_{\alpha}$\cite{Ezawa18,Ezawa18b,benalcazar2017quantized}. The SOTI phase, hosting corner modes, in modified Haldane model with IS i.e.,  without sub-lattice mass $M=0$, is described by $p_y=0.5$  and $p_x \ne 0.5$ (modulo unity) \cite{Wang21}. In our case with spin-full modified Haldane model, the SOTI phases are also expected to show half-integer quantization in presence of IS.  Under IS breaking $M\ne 0$, the quantization of $p_y$ is anticipated to deviate from the half-integer value that  we encounter below for various parameter regimes  while investigating the SOTI phases. We consider $\eta=0.25$ henceforth for all the remaining figures except  Fig.~\ref{fig:1} unless otherwise specified.

It  is, therefore, worth investigating 
the emergence of corner modes and their connection with the bulk-dipole moment. 
We show both the presence of mid-gap corner states, designated by turquoise-colored region, and finite spin-spectrum gap, bounded by red lines, in  the  Figs.~\ref{fig:1} (c) and (d), respectively for 
$V_{\rm SO} = 0.5$, and $2.5$ with $M=0,\eta=0.25$. We find  
half-integer quantization of $p_y$ within the regions where the spin-spectrum gap, computed for $\eta=0.25$, is finite except at $\phi=0,\pi$. 
Importantly, for small $V_{\rm SO}=0.5$ in Fig.~\ref{fig:1} (c), 
the QAHI and QSHI phases both can embed the SOTI phases as characterized by $p_y=0.5$. A part of FOTI phase is engulfed by vanishingly small spin-spectrum gap for $\eta=0.25$ and hence SOTI phase boundary is flanked with regard to the FOTI phase boundary. 
On the other hand, for relatively large $V_{\rm SO}=2.5$, the QAHI phase ceases to exist in the phase diagram leading to the SOTI phase out of the QSHI phases as shown in Fig.~\ref{fig:1} (d). However, the SOTI phases, embedded  in different FOTI phases, are not characteristically different as far as their spatial distribution and bulk-dipole moment are concerned. 
For $g=0$, the red line match with the boundary associated with turquoise-colored region in Fig.~\ref{fig:1} (d). One can refer to the mismatch for $g\ne 0$ as the apparent breakdown of generalized bulk-boundary correspondence in the $C_3$ symmetry-broken SOTI system that we leave for future study. However,  note that the present model breaks chiral symmetry and hence there is no restriction on the energy of the corner states.



\underline{\textit{Effect of different parameters:}}\\
\underline{\textit{1.~$M=0$ case:}}

Here we consider $M = 0, V_\text{SO} \neq 0, g \neq 0$ to investigate the mid-gap modes. These modes reside inside the SOTI phase as depicted in Fig.~\ref{fig:1} (c) and (d), in the presence (absence) of Zeeman field and SOC interaction (sub-lattice mass). For the cut over the weak [strong] bonds, referred to as cut 2 [cut 1],  edge modes cease [continue] to exist even when $|\eta|<0.5$ that is  depicted in Fig.~\ref{fig:2} (a) [(b)] under zigzag ribbon geometries \cite{Wang21}. One hence finds SOTI phase, hosting the localized mid-gap modes between the bulk gap, for cut 2 as shown in Fig.~\ref{fig:2} (c) where the energy spectrum of a nano-disc with cut 2 is demonstrated. The four mid-gap states inhabit only two corners at $(i,j)=(0,\pm L_y)$ while the remaining two corners at $(i,j)=(\pm L_x,0)$ are left unoccupied (see lower inset of Fig.~\ref{fig:2} (c)). The four-fold degeneracy of the mid-gap states can be achieved by tuning $g$ (see upper inset of Fig.~\ref{fig:2} (c)). Such a SOTI phase is characterized by $p_y$ which is found to be $0.5$ everywhere in $\phi$ except for $\phi=0,\pm \pi$ (see Fig.~\ref{fig:2} (d)). Note that $p_y$ remains quantized while $p_x$ behaves monotonically inside the blue shaded region where spin spectrum gap is finite. This refers to the fact that the WC lies at the middle of strong bond protecting the corner modes in cut 2 under IS.

The $C_3$ symmetry breaking with $|\eta|<0.5$ can gap out the edge states lying on the boundary of the nano-disc in cut 2 except at $(i,j)=(0,\pm L_y)$. In this geometry, the domain-wall only forms at corner  $(i,j)=(0,\pm L_y)$ where the two neighbouring cuts  over the weak bonds meet precisely at a strong bond \cite{Wang21}. The entire QSHI phase, as shown in Fig.~\ref{fig:1} (c) and (d), does not morph into the SOTI phase rather an interior part only can incubate SOTI phase. This is in stark contrast to the square lattice case where the entire QSHI phase becomes a quadrupolar insulator under $C_4$ symmetry and TRS breaking \cite{Dumitru19,Nag19,schindler2018higher}. Further notice that $p_y$
continues to exhibit the half-integer quantization irrespective of the energy of the mid-gap corner modes and the associated degeneracies (see Figs.~\ref{fig:2} (c) and ~\ref{fig:2} (d)).

To understand the behaviour of mid-gap corner states we need to focus on the three terms with $M, V_\text{SO}$ and $g$. We analyze the spin polarization of these states considering the fact that $g$ and $V_{\rm SO}$ break spin degeneracy while $M$ does not.   
When $g = 0$ and $M = 0$, because of the finite spin-orbit coupling, positive and negative energy mid-gap states are spin polarized with a gap between them. We refer to these modes as spin-polarized pairs.  We depict such pairs for $g \ll V_{\rm SO}$ in Fig.~\ref{fig:2} (c) where blue (red) denotes the spin-up (down) polarization.
 Thereafter, the gap between these spin polarized pairs can be manipulated by the varying $g$. To be precise,   
 the intra-pair gap does not change rather inter-pair gap  is only modified. This 
 leads to the two pairs of zero-energy corner modes while each pair is comprised of two opposite spin-polarized corner modes under a suitable choice of parameters. 
 

\underline{\textit{2.~$g=0$ case:}}
We now turn our attention to the SOTI phase that is embedded in the underlying QSHI phase  when $|\eta|<0.5$ as shown in Fig.~\ref{fig:1} (d).
We consider here, $M\neq 0,V_{\rm {SO}} \neq 0$, and $g = 0$ case where we find that the degeneracy of the mid-gap states at non-zero energy is lifted due to the finite sub-lattice mass  while the bulk gap is controlled by $V_{\rm SO}$. With changing $M$, two out of four mid-gap states can be brought to zero-energy and the remaining two move further away from zero-energy, towards the bulk levels. This refers to the fact that four-fold degeneracy of the corner states at zero-energy reduces to two-fold. The four non-degenerate mid-gap states are depicted in  Fig.~\ref{fig:3} (a). Figure \ref{fig:3} (c) demonstrates another instance where the QSHI hosts doubly degenerate zero-energy modes along with two finite-energy non-degenerate corner states. The band dispersion for the above situation in cut 2 ribbon geometry clearly shows a gap as evident from  Fig.~\ref{fig:3} (b). One can find that for positive (negative) values of $M$, $p_y$ remains close to unity (zero) [Fig.~\ref{fig:3} (d)]. This is a clear indication of corner modes in SOTI phase that is not characterized by the bulk-dipole moment.

We now investigate the spin polarization of these mid-gap pairs. 
Extending the prior understanding for
$M= 0$, we  here find that $M\ne 0$ causes the intra-pair gap for a given spin-polarized pair. On the other hand, $V_\text{SO}$ is simultaneously responsible for the inter-pair as well as intra-pair gaps. 
There exists a competition between $V_\text{SO}$ and $M$ such that the intra- and inter-pair gaps both can be tuned. For example, the one mode  inside a given spin-polarized pair, residing at positive energy, moves further away from its counterpart within the above pair and eventually gets closer to the mode with negative energy coming from the opposite spin-polarized pair.  Therefore, by tuning $M$, one corner mode from each spin-polarized pair can come arbitrarily close to zero-energy while the remaining corner modes with opposite spin polarization reside inside the bulk gap away from zero-energy.
This allows us to have  a two-fold degenerate zero-energy corner mode out of four mid-gap modes which are also evident from Fig.~\ref{fig:3} (c). As a result, one can obtain one pair of zero-energy corner modes with opposite spin polarization. This is in contrast to the previous case with $M=0$ where we find two pairs of zero-energy corner modes.



\underline{\textit{3.~$V_{\rm SO}=0$ case:}}
Here we find that depending on the relative sign and magnitudes of $M$ and $g$ two mid gap states move into the bulk while remaining two mid gap states stays at zero energy. The evolution of mid-gap states is demonstrated in Figs. \ref{fig:4} (a), (b) by varying $g$ and keeping $M$ fixed at $0.5$ such that $|g|<M$. We repeat Figs. \ref{fig:4} (a),  and (b)
in  Figs. \ref{fig:5} (a), and (b), respectively, for $|g|\ge M$. Interestingly, for $M = -g$,  we find that two mid-gap states out of four mid-gap states can be constrained at zero-energy. Importantly, the bulk-gap around zero-energy reduces with increasing $|g|$.  The continuous bulk bands, observed for $|g|<M$, become gapped out at finite energies within which the mid-gap corner state can be present for $|g|>M$.  The individual [combined] localizations for the mid-gap states are depicted in Figs. \ref{fig:4} (c1-c4) [Fig. \ref{fig:4} (d1)]. Similarly, the local density of states for the mid-gap modes is depicted in  Fig. \ref{fig:5} (c1) associated with Figs. \ref{fig:5} (a). 
This is characteristically different from the quadrupolar insulator where each of the mid-gap states populate more than a single corner of a 2D square lattice \cite{Nag19}. The ribbon geometry clearly indicates the gapped band structure in Figs. \ref{fig:4} (b) and (d). The polarization $p_y$ changes with $M$ demonstrating the breakdown of topological characterization for IS broken case irrespective of the filling (see Fig.~\ref{fig:5} (d1)). We show the gapped band structure under ribbon geometry for cut 2 in  Fig. \ref{fig:4} (d2), corresponding to Fig. \ref{fig:4} (b), that  refers to the existence of SOTI phase. Likewise, we repeat the above in Figs.~\ref{fig:5} (c2) and (d2) for Figs.~\ref{fig:5} (a) and (b), respectively, indicating the absence of edge modes in cut 2.

We now analyze the spin polarization of the mid-gap states with respect to $g$ and $M$. In this case, the pair at finite energy is composed of corner modes with opposite spin-polarization (see Fig.~\ref{fig:4} (a)). With increasing $|g|$, the intra-pair increases leading to the disappearance of one of the mid-gap states from each pair, having opposite spin-polarization, into the bulk bands.   When $|g|=|M|$, the remaining
two opposite spin-polarized corner modes, coming from two different mid-gap pairs, get pinned down at zero-energy. Therefore, one can find one pair with opposite spin polarization similar to the earlier case with $g=0$. However, the basic difference here is that we reduce the number of  mid-gap corner modes to two  from four as obtained for $g=0$ (see Figs.~\ref{fig:3} (a), (c), Fig.~\ref{fig:4} (b) and Fig.~\ref{fig:5} (a)).

	
\underline{\textit{4.~
General case:}}
We demonstrate the mid-gap corner modes in the SOTI
phase in Figs.~  \ref{fig:6} (a) and (c) for nano-disc and ribbon geometry band structure, respectively. The SOTI phase examined here is encapsulated within the underlying QSHI phase as depicted  in Fig.~  \ref{fig:1} (b). A close inspection suggests that the SOTI  phase,  is not embedded in the entire QSHI phase rather an interior part of it for $M,g\ll V_{\rm SO}$, $\phi < \pi/4$ as qualitatively estimated from the effective band-gaps for $|\eta| < 0.5$. This is depicted as an inset in  Fig.~  \ref{fig:6} (a) where 
mid-gap corner states are absent when the parameters
are chosen from an exterior part of the underlying QSHI phase. 
A similar phenomenon is observed in Fig.~\ref{fig:1} (c) and (d). However,  the determination of the exact phase boundary for the SOTI phase in $M-\phi-g-V_{\rm SO}$ parameter space is beyond the scope of the present study as the appropriate topological invariant is yet to be determined for IS broken case.  
We now extend our analysis to QAHI phase as shown in Fig.~~\ref{fig:1} (c) with $V_{\rm SO}=0.5$ and $M=0$. We find there four zero-energy mid-gap corner modes as a signature of inversion symmetric SOTI phase that emerges from QAHI phase when $\eta=0.25$  (see Fig.~ \ref{fig:6} (b)). The corresponding gapped ribbon geometry clearly suggests the absence of the edge modes in the SOTI phase  as shown in Fig.~ \ref{fig:6} (d). We further examine the  case with $V_{\rm SO}=0.5$ and $M\ne 0$ from Fig.~\ref{fig:1} (a) where two mid-gap corner modes, depicted as an inset in Fig.~ \ref{fig:6} (b),  bear the signature of inversion broken SOTI phase embedded in  QAHI phase. 


\underline{\textit{Discussion and conclusion:}}

We find that a hybrid symmetry operator defined as $ {\mathcal C}=\sigma_2 \tau_1$ yields ${\mathcal C} H(k_x,k_y,\phi){\mathcal C}^{-1}= -H(k_x,-k_y,\pi-\phi)$. It also maps $A\sigma \to B\bar{\sigma}$ ($\bar{\sigma}=-\sigma, \sigma =\uparrow, \downarrow$). This symmetry operator
dictates the composite topological character of the both the spin-sectors and allows to have HOTI phases when the parent FOTI phase is QSHI or QAHI phases. It prohibits the QASHI to generate HOTI phases as both the spin-sectors are not topological. Interestingly, when $V_{\rm SO}=0$ and $M=-g$, the terms  associated with $c_{A\sigma}^{\dagger} c_{A\sigma}$ and $c_{B\bar{\sigma}}^{\dagger} c_{B\bar{\sigma}}$ in the Hamiltonian (Eq.~(\ref{main:ham})) are identical as $M$ and $g$ appear in identical sign.  Such reciprocity can be understood from the structure of the symmetry operator ${\mathcal C}$. It is noteworthy that the above choice of  parameter set yields a pair of zero-energy mid-gap corner states with opposite spin polarization as shown in Fig.~\ref{fig:5} (a). Therefore, one can comment that the hybrid symmetry effectively constrains the zero-energy mid-gap states having opposite spin components as evident for even and odd pairs in Figs.~\ref{fig:2} (c) and ~\ref{fig:3} (c), respectively.


The FOTI phases for $\eta=0.5$ might be useful to understand the emergence of SOTI phases for $0<\eta
<0.5$. Note that Hamiltonians at the ${\bf M}$-points need to be gapped out to host the SOTI phase. 
The energies at ${\bf M}$ points for the spin up and spin down sub-blocks are given by $E^{\pm}_{\uparrow}=-V_\text{SO} \sin \phi/3  + g  \pm  \sqrt{(2\eta t_{1} - t_{1})^2+M^2}$ and $E^{\pm}_{\downarrow}=V_\text{SO} \sin \phi/3  - g  \pm  \sqrt{(2\eta t_{1} - t_{1})^2+M^2}$. However, one can comment that the diagonal blocks are always individually gapped out for $\eta <0.5$ irrespective of the values of $V_\text{SO}$ and $g$ even if $M=0$. Interestingly, complete $4$-level Hamiltonian is also gapped out for $g=V_\text{SO} \sin \phi/3$ that is when the energies for spin-up and down sectors match each other. Interestingly, for $M=-g\ne 0$ and $V_\text{SO}=0$, there always exists a gap within the individual spin sector 
as well as between two different spin sectors. A similar gap structure   can be observed for $M,V_\text{SO}\ne0$, and $g=0$ case as well. 
Therefore, the separation between the spin up and down mid-gap corner modes can be controlled by $g$, $\phi$ and $V_{\rm SO}$ while $M$ acts identically for all spin sectors. Such a complicated interplay is clearly observed in Figs.~\ref{fig:2}, \ref{fig:3}, \ref{fig:4}, \ref{fig:5} where we find even or odd pairs of zero-energy modes under suitable parameters. We would like to comment that there could be instances when  any two out of the four energy levels $E^{\pm}_{\uparrow,\downarrow}$ become degenerate. This would then cause the collapse of the SOTI phase as ${\bf M}$-points no longer remain gapped out.


Another important point to note is that apart from the band gap, a finite spin-spectrum gap is a necessary criteria for any topological phase irrespective of its order. The spin-spectrum gap, computed for $|\eta|<0.5$  ($|\eta|=1$) can be referred to as SO (FO) spin-spectrum gap. To be precise, in our case FO and SO spin-spectrum gaps both  have to be finite in order to host mid-gap corner modes while edge modes can be present only for finite FO spin-spectrum gap (see Fig. \ref{fig:1} (c)). On the other hand, the half-integer quantization of $p_y$, characterizing the inversion symmetric SOTI phase, is assured when there exists  the SO spin-spectrum gap (see Figs. \ref{fig:1} (c), and (d)). Interestingly, the $p_y$ continues to show half-integer quantization even though the mid-gap states cease to exist (see Figs.~\ref{fig:1} (c), and (d) ). Therefore, there is an open question why $p_y$ continues to show half-integer quantization even if there exist no mid-gap SOTI corner states as well as in absence of the FO spin spectrum gap.

 To conclude, we analyzed the fate of eight-fold phases \cite{Saha21} of extended Haldane model in the anisotropic limit and found that  QSHI and QAHI phases can give rise to SOTI phases while IS is preserved. QASHI phase never give rise to any SOTI phases as found in our analysis. We show how to manipulate the zero energy corner modes from four-fold degeneracy to two-fold degeneracy by suitable application of sub-lattice mass, Zeeman field and SOC interaction (See Figs 2, 3, 4, 5). The bulk-dipole moment is able to characterize the SOTI phase as long as IS is preserved. We showed that a hybrid symmetry allows the QSHI and QAHI phase to embed the SOTI phases and prohibits the QASHI phases to embedded any HOTI phases. Remarkably a finite spin-spectrum gap continues to be a necessary condition for the existence of SOTI phases as well. The experimental realization of SOC interaction  \cite{lin2011spin,huang2016experimental,meng2016experimental} and Haldane model \cite{jotzu2014experimental,liu2018generalized} allow our model to become experimentally viable. The disordered and Floquet extension of these phases can be the open future directions.

\section{Acknowledgement} 
SS acknowledges SAMKHYA (High-Performance Computing Facility provided by the Institute of Physics, Bhubaneswar) for the numerical computation.

\bibliography{apssamp}{}
\end{document}